\documentclass{llncs}
\usepackage{amsmath}
\usepackage{subfig}

\usepackage{verbatim}
\usepackage{todonotes}
\usepackage{url}

\author{James Stovold \orcidID{0000-0002-0708-2630}} 
\institute{Department of Computer Science \\ Swansea University \\ Swansea, Wales \\ SA1 8EN \\ \url{j.h.stovold@swansea.ac.uk}} 
\title{Trainable Associative Memory Neural Networks in a Quantum-Dot Cellular Automata}

\begin{document}
\maketitle

\begin{abstract} 

Quantum-dot cellular automata (QCAs) offer a diffusive computing paradigm with picosecond 
transmission speed, making them an ideal candidate for moving diffusive computing to 
real-world applications. By implementing a trainable associative memory neural network into 
this substrate, we demonstrate that high-speed, high-density associative memory is feasible 
through QCAs. The presented design occupies $415\text{nm}^2$ per neuron, which translates to 
circa $240 \text{ billion neurons/cm}^2$, or $28\text{GB/cm}^2$ of memory storage, offering a real 
possibility for large-scale associative memory circuits. Results are presented from 
simulation, demonstrating correct working behaviour of the associative memory in single 
neurons, two-neuron and four-neuron arrays.

\end{abstract}

\section{Introduction}
\label{sec:intro}

 % benefits of using qca over "traditional" methods
 % why are we building CMMs?
 % include more self-citations

Quantum-dot cellular automata (QCAs) are a transistor-less computing paradigm that uses a 
bistable arrangement of electrons within quantum dots to represent binary 
values~\cite{gin_alternativegeometryquantum,lent_bistablesaturationcoupled}. The QCA 
paradigm~\cite{tougaw_dynamicbehaviorquantum} offers a route towards real-world application 
for computational constructs that have been developed in naturally diffusive systems, such as 
reaction--diffusion 
chemistry~\cite{stovold_simulatingneuronsreaction,stovold_reactiondiffusionchemistry,stovold_associativememoryreaction,adamatzky_collisionbased}, 
slime mould~\cite{shirakawa_onsimultaneousconstruction}, and billiard ball 
computing~\cite{durandlose_computinginsidebilliard}.

This paper presents a QCA-based associative memory neural network, in the form of a 
correlation matrix memory (CMM). A CMM is a weightless neural network that associates 
stimulus--response pairs~\cite{kohonen_correlationmatrixmemories}. CMMs have been 
successfully applied to a wide range of problems, including address matching for the UK Post 
Office~\cite{lomas_postoffice}, matching patterns in chemical 
structures~\cite{turner_neuralrelaxationtechnique}, and image 
processing~\cite{austin_distributedassociativememory}.

QCA-based neural networks, such as those presented in this paper, have the potential for 
high-speed pattern matching on a scale that exceeds that of the human brain by multiple 
orders of magnitude. The human brain has a neuronal density of approx.~$720 \pm 69 \text{ 
million/cm}^3$~\cite{lange_cellnumbercell}, whereas the proposed neural network occupies 
approx.~$415\text{nm}^2$ per neuron, giving a potential neuronal density of approx.~$240 
\text{ billion/cm}^2$.

This paper is structured as follows: the preliminary background of QCAs and CMMs are 
presented in section~\ref{sec:prelim}; section~\ref{sec:implementation} presents the 
implementation of CMMs in QCA, starting with a single neuron, moving up to arrays of 2 and 4 
neurons; finally, section~\ref{sec:discussion} concludes the paper. Due to space constraints 
the simulation results for four-neuron arrays and horizontal two-neuron arrays are provided 
in the appendix.

\section{Preliminaries}
\label{sec:prelim}
\subsection{Quantum-Dot Cellular Automata}
\label{sec:qca}

\begin{figure}[h!]
 \centering 
 \includegraphics[width=0.8\linewidth]{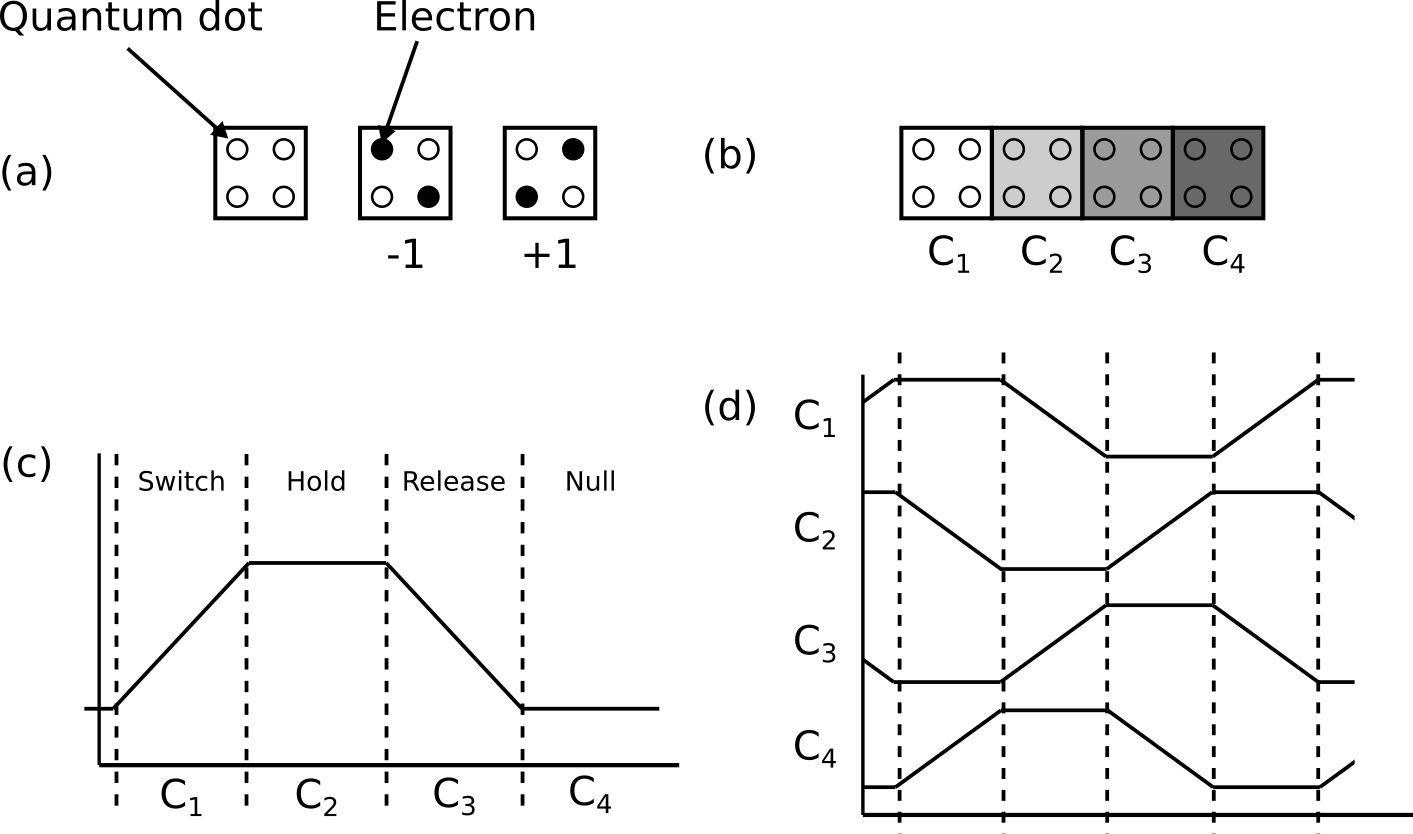}
 \caption[]{ (a) The QCA cell consists of four quantum dots and two electrons. The two 
electrons can tunnel between quantum dots. The two polarisations shown ($-1$ and $+1$) 
represent logical 0 and logical 1, respectively. (b) QCA cells can be split into different 
clock cycles, depicted by varying colours of the cell. (c) The clock cycle for a single QCA 
cell: when the clock cycle is at `hold' the electrons are not able to tunnel between dots. 
(d) By clocking adjacent cells in turn, information propagates between cells in the clocking 
direction. }
 \label{fig:qca} 
\end{figure}

Quantum-dot cellular automata (QCAs) are a transistor-less computing 
paradigm~\cite{lent_bistablesaturationcoupled}. A QCA consists of a regular array of `QCA 
cells', each consisting of four quantum dots and two electrons (see fig.~\ref{fig:qca}(a)). 
The cell is arranged such that the electrons are free to tunnel between quantum dots within 
an individual cell, but not to neighbouring cells. An electron moves based on the Coulomb 
repulsion forces of the other electron in the cell and of those in neighbouring cells. As 
such, the cell has two stable states with $-1$ and $+1$ polarisation. By treating these 
stable polarisation states as logical 0 and 1 respectively, logical circuits can be 
constructed~\cite{tougaw_logicaldevicesimplemented}.

QCA circuits can be clocked by varying the tunneling potential between each quantum 
dot~\cite{lent_devicearchitecturecomputing}. The tunneling potential determines whether the 
electron can tunnel from one dot to another. As depicted in fig.~\ref{fig:qca}(c) and (d), by 
varying this potential over time, the electron can be held in place or be allowed to switch 
freely. This provides a mechanism through which directional propagation of state can be 
achieved. Fig.~\ref{fig:qca}(b) shows how this varying clock signal is depicted in this 
paper. In QCADesigner~\cite{walus_qcadesigner}---the simulator used in 
section~\ref{sec:implementation}---these are represented by the colours green, pink, blue, 
and white.

\begin{figure}[h!]
 \centering
 \includegraphics[width=0.6\linewidth]{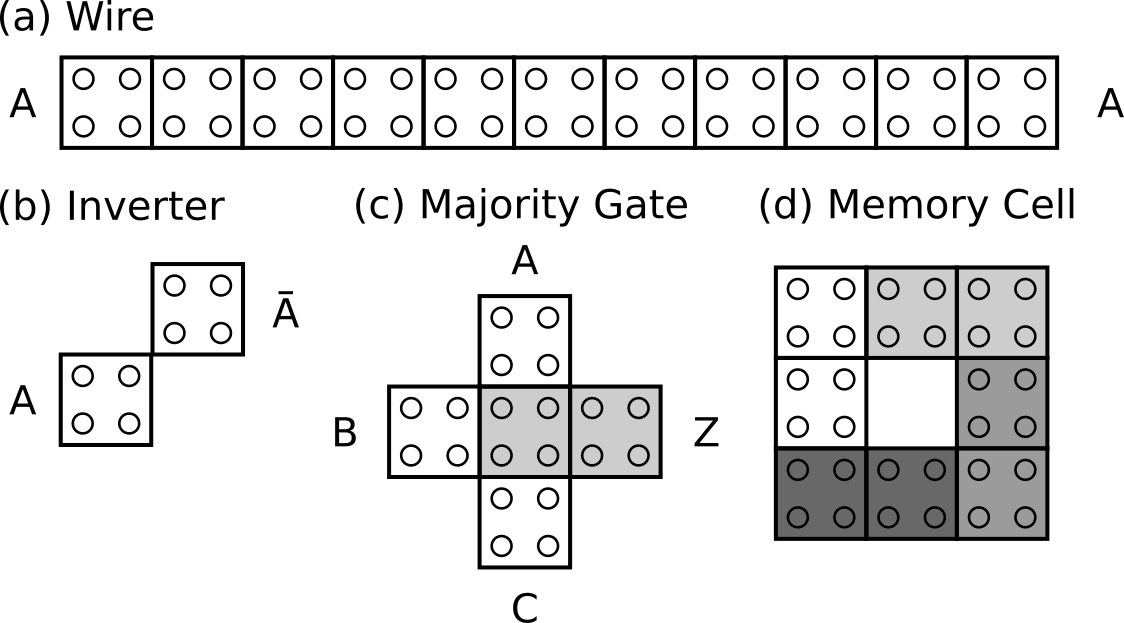}
 \caption[]{Various logical circuits can be constructed in QCA cells, including (a) wires, 
(b) inverters, (c) majority gates and (d) memory cells.}
 \label{fig:qca_logic}
\end{figure}

The basic building blocks of logical circuits were constructed by Tougaw and 
Lent~\cite{tougaw_logicaldevicesimplemented}, who demonstrated that signals can propagate 
along a `wire' of QCA cells \cite{lent_linesinteractingquantum} (see 
fig.~\ref{fig:qca_logic}(a)) and that logical devices can be constructed, including logical 
inverters (b) and a majority gate (c). The three-input majority gate can be used to implement a 
two-input AND or OR gate, depending on the value of the third input (logical 0 for AND, 1 for 
OR).

% clocking devices

Further to the logic gates developed by Tougaw and 
Lent~\cite{tougaw_logicaldevicesimplemented}, memory cells have been developed using 
loop-based~\cite{berzon_memorydesignqcas} (fig.~\ref{fig:qca_logic}(d)) and line-based 
architectures~\cite{vankamamidi_linebasedparallel,taskin_improvinglinebased}. QCA memory 
designs include D flip-flops~\cite{hashemi_newrobustqca,zoka_novelrisingedge} and RAM 
devices~\cite{walus_ramdesignusing,ottavi_designqcamemory,ottavi_novelmemorydesigns}.

% why build neural nets into QCAs?

While artificial neural networks and associative memory have been designed for the quantum 
computing 
paradigm~\cite{ezhov_quantumneuralnetworks,perus_neuralnetworksbasis,schuld_introductionquantummachine,ventura_quantumassociativegrover,ventura_quantumassociativememory}, 
there is no associative memory for the QCA paradigm, only the analogue cellular neural 
network proposed by Toth et al.~\cite{toth_quantumcellularneural} which does not consider 
memory formation.

%The implementation presented here uses only 120 cells per bit, and is able to write to memory 
%in O(1) time. For larger implementations (i.e.~greater than 2 neurons), we have to introduce 
%a clocking delay, which reduces the write speed to O(n), see section~\ref{sec:four-neuron} 
%for details. This is the subject of further study.

 % transistor-less computing paradigm
 % bistable cells consisting of four quantum dots with two electrons
 % the Coulomb repulsion of the electrons results in two stable states with +1 and -1 polarisation
 % by treating these stable polarisation states as logical 1 and 0, we are able to construct 
 %logical devices

 % 

 % what are quantum dots
 % what is the QDCA / QCA
 % basic logic gates etc. in QCA
 % why is this a good option compared with RDC?

\subsection{Correlation Matrix Memories}
\label{sec:cmm}

 % what is associative memory
 % what are CMMs
 % what are the main benefits of using CMMs over other architectures
 % training 
 % recall
 % thresholding 

\begin{comment}
 associative memory from Palm -- how neurons start to form memory through basic Hebbian learning

correlation matrix memories are a form of associative memory. associative memory associates 
stimuli with responses. this can be implemented as a fully-connected, binary-weighted, 
two-layer artificial neural network. 

one of the key benefits of correlation matrix memories is their ability to generalise noisy 
inputs. this means if we were to present a noisy version of a stimulus, we would still 
retrieve the correct response.

the correlation matrix memory was developed from Willshaw's `associative net' by Kohonen and 
popularised by Austin et al

the cmm is best thought of as the weight matrix of a fully-connected, binary-weighted, 
two-layer artificial neural network (one input-layer, one output-layer). the neural network 
depicted in fig.~\ref{fig:cmm_nn} would be represented by the CMM, $\mathcal{M}$, with $k$ 
input--output pairs $\mathcal{I}$ and $\mathcal{O}$:

\end{comment}
% ==

Associative memory is a form of memory that associates stimulus--response pairs. In 
biological systems, the memory results from temporal associations that emerge between two 
sets of interacting 
neurons~\cite{palm_onassociativememory,palm_towardstheorycellassemblies,willshaw_nonholographic}. 
As the stimulus is presented to the first set of neurons, the response of those neurons is 
sent to the second set of neurons, which subsequently respond. As this second response is 
similar each time, the network of neurons can be said to have learnt the association between 
the two signals.

\begin{figure}[h!]
 \centering
 \subfloat[Basic CMM architecture (left), where $\mathcal{M}$ represents the matrix of binary
weights, $\mathcal{I}$ and $\mathcal{O}$ represent the input--output pair corresponding to
the neurons $a,b,c$ and $d,e,f$ in (b), respectively. \label{fig:cmm:maths}]{ \centering
\parbox{0.5\linewidth}{ %
 \begin{displaymath}
 \begin{array}{lc|cr}
 \begin{matrix}
   \begin{pmatrix}
     ~ \\
         \mathcal{I} \\
         ~
    \end{pmatrix}
   &
        \begin{pmatrix}
         ~ & ~ & ~ \\
         ~ & \mathcal{M} & ~ \\
         ~ & ~ & ~
        \end{pmatrix}
   \\
        ~ \\
   &
 \begin{pmatrix}
         ~ & ~\mathcal{O} & ~~
        \end{pmatrix}
  \end{matrix}
 & ~~ & ~~ &
  \begin{matrix}
    \begin{pmatrix}
        a \\
        b \\
        c
        \end{pmatrix}
  &
        \begin{pmatrix}
      ad & ae & af \\
          bd & be & bf \\
          cd & ce & cf
        \end{pmatrix}
  \\
        ~ \\
  &
        \begin{pmatrix}
          ~d & ~e & ~f~
        \end{pmatrix}
  \end{matrix}
 \end{array}
 \end{displaymath} \vspace{2em} }
} ~
 \subfloat[CMM associative memory neural network. The binary weights between the two layers
are represented by the matrix $\mathcal{M}$ in (a). \label{fig:cmm:nn}] { \centering %
\parbox{0.45\linewidth} {
\centering
  \includegraphics[width=\linewidth]{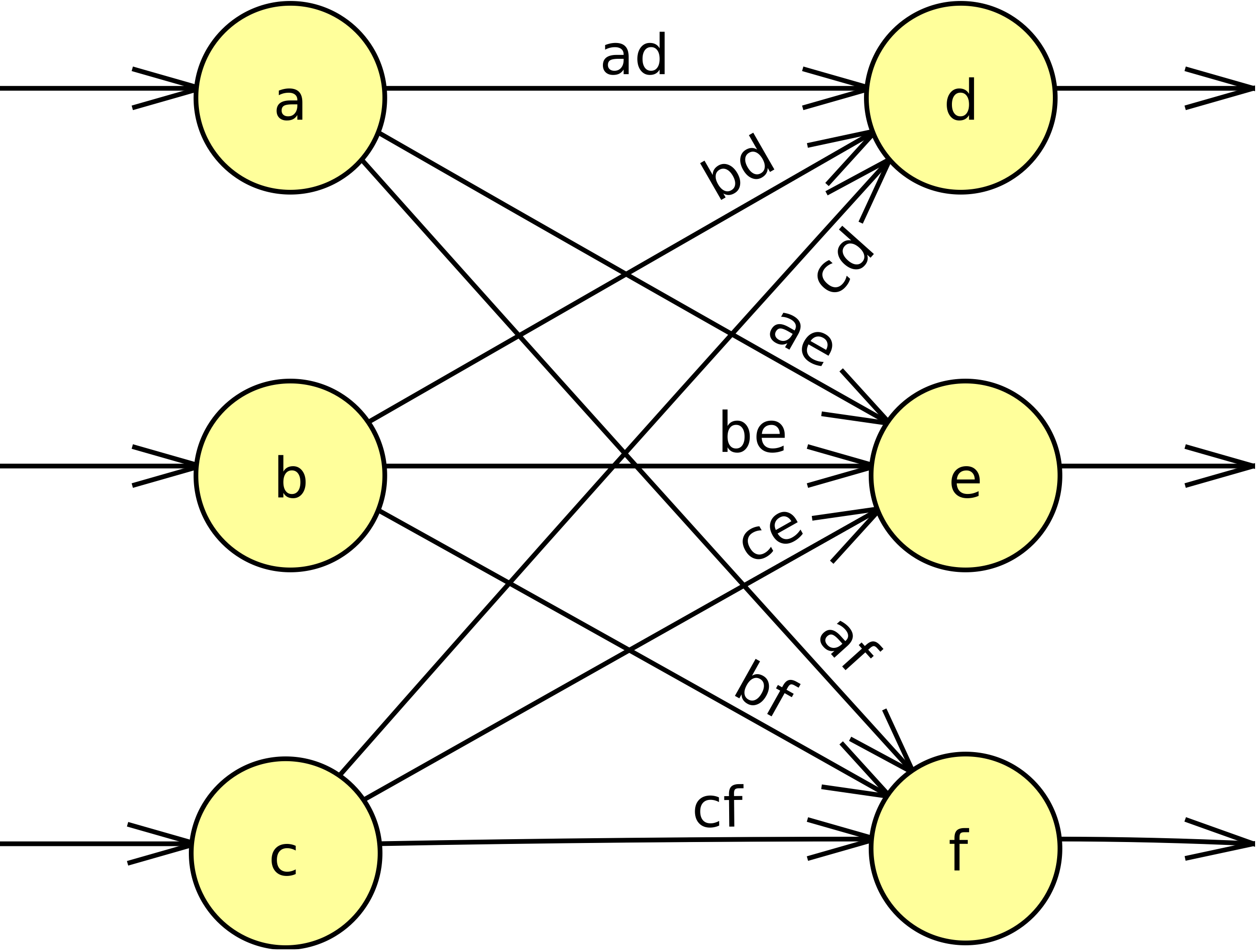} }
}
 \caption[]{}
 \label{fig:cmm}
\end{figure}

The CMM (Correlation Matrix Memory) is a matrix-based representation of this 
process~\cite{kohonen_correlationmatrixmemories}. The two sets of interacting neurons can be 
represented as a fully-connected, two-layer artificial neural network (one input layer, one 
output layer; see fig.~\ref{fig:cmm:nn}). Associative memory can be formed by presenting 
stimulus--response pairs of binary vectors, training the connections between the two layers. 
The CMM is the binary weight matrix that connects these two layers.

For example, the network in fig.~\ref{fig:cmm:nn} would be represented by the CMM, 
$\mathcal{M}$, with $k$ input--output pairs $\mathcal{I}$ and $\mathcal{O}$ in 
fig.~\ref{fig:cmm:maths}.

Before training, the initial matrix $\mathcal{M}$ is filled with zeros (as there are no
associations stored in the network). As the $k$ binary-valued input--output pairs are
presented to the network, associations are built up in the matrix $\mathcal{M}$. These
associations are stored as 1s in the matrix, corresponding to coincident 1s in both input and
output vectors. For example, in fig.~\ref{fig:cmm:maths}, if $a$ and $e$ were both 1 then
$ae$ would be set to 1 after training.

% recall details

\begin{comment}

to recall stored patterns, the original stimulus (or some noisy version thereof) is 
re-presented to the network, and the network will respond with the closest stored pattern:

\end{comment}

To recall, an input pattern $\mathcal{I}_r$ is presented to the 
network:~\cite{haykin_neuralnetworks}
 \begin{displaymath}
   \mathcal{O} = \mathcal{MI}_r
 \end{displaymath}
 \noindent where $\mathcal{I}_r$ is the input pattern, and $\mathcal{O}$ is the output from
the network trained with associations stored in $\mathcal{M}$. The desired output pattern,
$\mathcal{O}_r$ is currently combined with noise from the other patterns stored in the
network, $e_r$, hence:
 \begin{displaymath}
  \begin{array}{l}
   \mathcal{O} = \mathcal{O}_r + e_r \\
   e_r = \sum\limits_{\substack{j=1\\j\neq r}}^{k} (\mathcal{I}^{T}_j \mathcal{I}_r)\,\mathcal{O}_j
  \end{array}
 \end{displaymath}
 \noindent Thresholding the output vector $\mathcal{O}$ leaves the desired output vector
$\mathcal{O}_r$. Different thresholding strategies offer different advantages in terms of the
ability of the network to generalise from noisy or incomplete patterns to a correct
output~\cite{austin_distributedassociativememory}.

The ability of the network to generalise noisy inputs suggests a range of applications in 
real-world environments, where a noisy signal is far more common than a clean signal. Example 
applications areas include stock market 
prediction~\cite{kustrin_applicationcorrelationmemory}, fault 
detection~\cite{liang_mininglargeengineering}, address matching~\cite{lomas_postoffice}, 
associating signals in 
robots~\cite{stovold_distributedcognitionbasis,stovold_cognitivelyinspiredhomeostatic} and 
image processing~\cite{austin_distributedassociativememory}.

\section{Implementation}
\label{sec:implementation}

 % simulation approach (using QCADesigner)
 % are there any issues with this approach?
 % how well do they transfer across to real QCAs?

This section presents the QCA-based implementation of CMMs. Throughout this section, the 
correct behaviour of the designs has been demonstrated through a simulator called 
QCADesigner~\cite{walus_qcadesigner}. 

\subsection{Individual Neurons}
 
 % basic design of individual neuron
 % implementation in QCADesigner
 % testing individual neuron (training)
 % recall?

The essence of the CMM neuron is to detect and store coincident input bits between binary 
arrays (see section~\ref{sec:cmm}). Coincidence detection is achievable using a logical AND 
gate, which is achievable using a majority gate with one of the inputs fixed to logical 0 
($-1$ polarisation)~\cite{tougaw_logicaldevicesimplemented}. Combining this AND gate with a 
memory cell is enough to detect and store the coincidence of logical 1 inputs.

\begin{figure}[h!]
 \centering
 \includegraphics[width=0.45\linewidth]{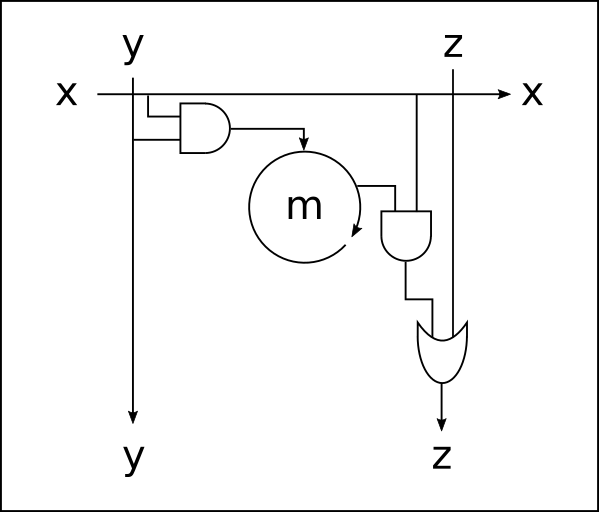}~
 \includegraphics[width=0.4\linewidth]{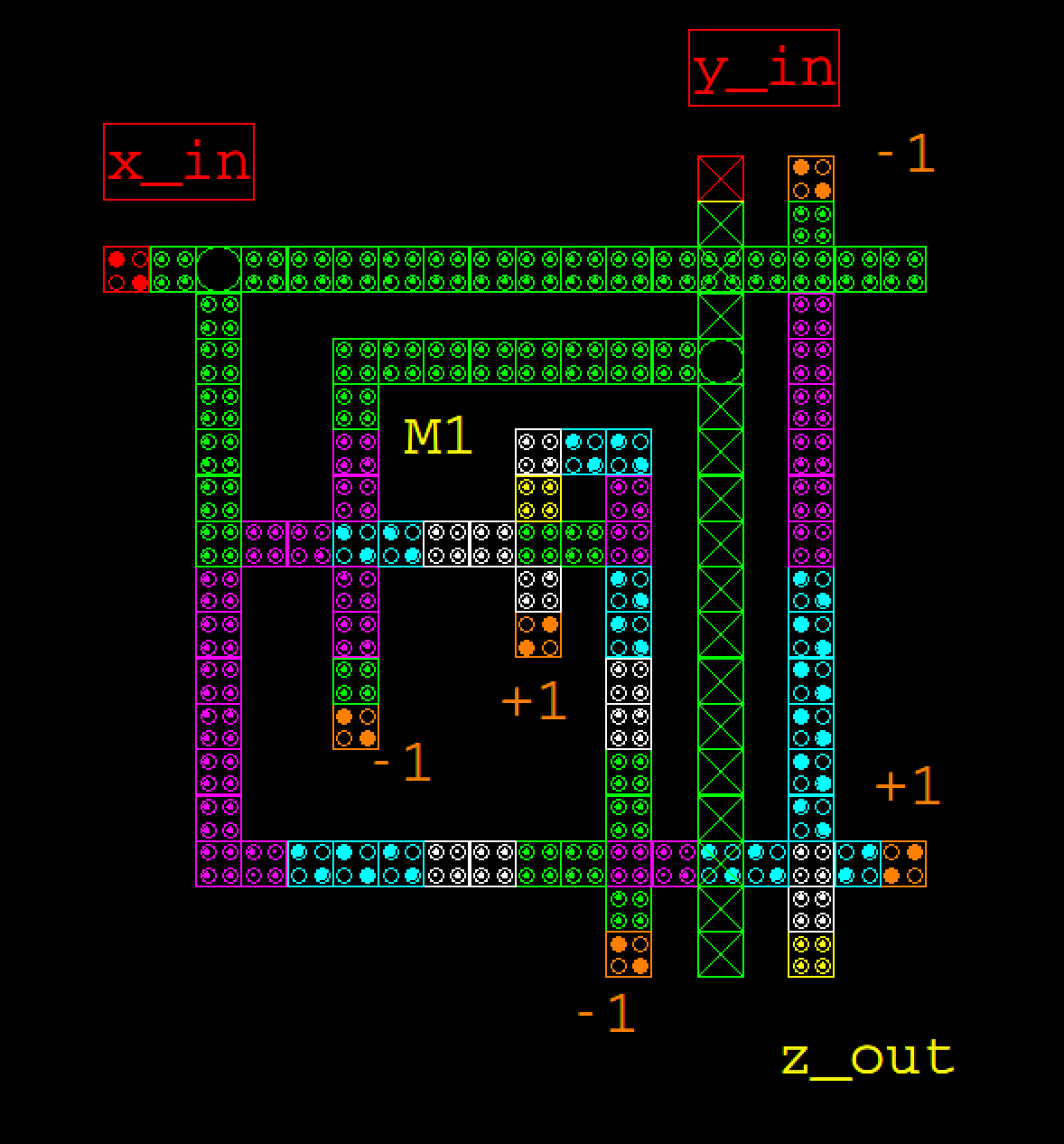}
 \caption[]{ Left: logical design of a single CMM neuron. Right: QCA design of a single CMM 
neuron. Two individual bits are presented to the neuron through \texttt{x\_in} and 
\texttt{y\_in}. The memory cell is then set to $M = x \wedge y$. The value of the memory cell 
is then outputted to the lower AND gate. This gate controls the recall procedure---the value 
of the \texttt{M1} is only passed to \texttt{z\_out} when \texttt{x\_in} is 1. }
 \label{fig:single_neuron:design}
\end{figure}

Fig.~\ref{fig:single_neuron:design} shows the logical design and implementation for a single 
CMM neuron in QCADesigner. Two single-bit inputs (\texttt{x\_in} and \texttt{y\_in}) 
represent single bits in the two binary vectors presented to the CMM during the training 
phase of the network. Note that the \texttt{x\_in} and \texttt{y\_in} wires are on different 
levels to the main neuron, to prevent interference between input signals, although this is 
not required it makes simulating the design 
simpler~\cite{graunke_implementationcrossbarnetwork}. To recall the value from the neuron, 
the \texttt{x\_in} input is set to 1, which allows the AND gate below the memory cell to pass 
the value of the memory cell to \texttt{z\_out}.

\begin{figure}[h!]
 \centering
 \includegraphics[width=0.7\linewidth]{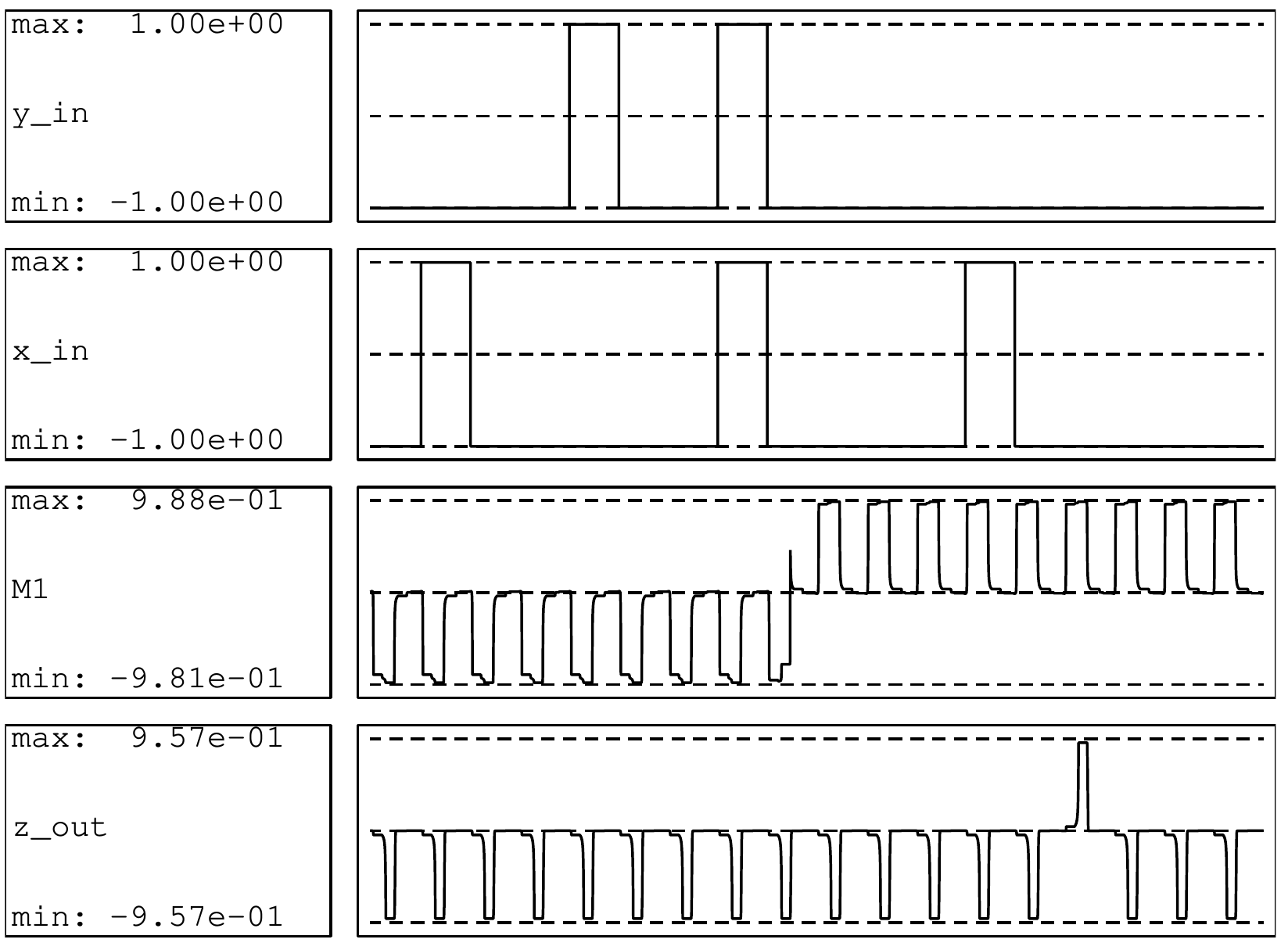}
 \caption[]{Simulation results for a single CMM neuron. The CMM neuron is trained through the 
\texttt{x\_in} and \texttt{y\_in} inputs, associating the two input values by setting the 
memory cell \texttt{M1} if both inputs are simultaneously 1. The associated value is recalled 
by setting \texttt{x\_in} to 1, at which point the output \texttt{z\_out} is set to the value 
of \texttt{M1}.}
 \label{fig:single_neuron:trace}
\end{figure}

Fig.~\ref{fig:single_neuron:trace} shows the signal traces from simulating the neuron 
depicted in fig.~\ref{fig:single_neuron:design}. The memory cell within the CMM neuron 
(\texttt{M1}) is only set to 1 when both inputs are 1, and the output from the neuron 
(\texttt{z\_out}) is set to 1 when the memory cell is set and \texttt{x\_in} is set to 1, 
demonstrating that the stored value can be recalled.

\subsection{2-neuron array}
\label{sec:two-neuron}

 % connecting two neurons together (V and H)
 % issues encountered 
 % - multi-layer CA
 % testing 2-neuron array

The logical next step in developing a usable QCA-based CMM is to connect multiple neurons 
together. Connecting multiple neurons in a vertical and horizontal arrangement is not 
necessarily trivial: Coloumb interference between neurons, crossing lines of cells, and long 
chains of cells in a wire can all introduce unexpected effects.

\begin{figure}[h!]
 \centering 
 \subfloat[Horizontal arrangement of two CMM neurons. The \texttt{x\_in} input is passed 
across from the first neuron to the second, and two \texttt{y\_in} inputs are provided, one 
for each neuron. \label{fig:two-neuron:design:hoz}] {\centering
\parbox{0.8\linewidth}{ \centering
 \includegraphics[width=0.65\linewidth]{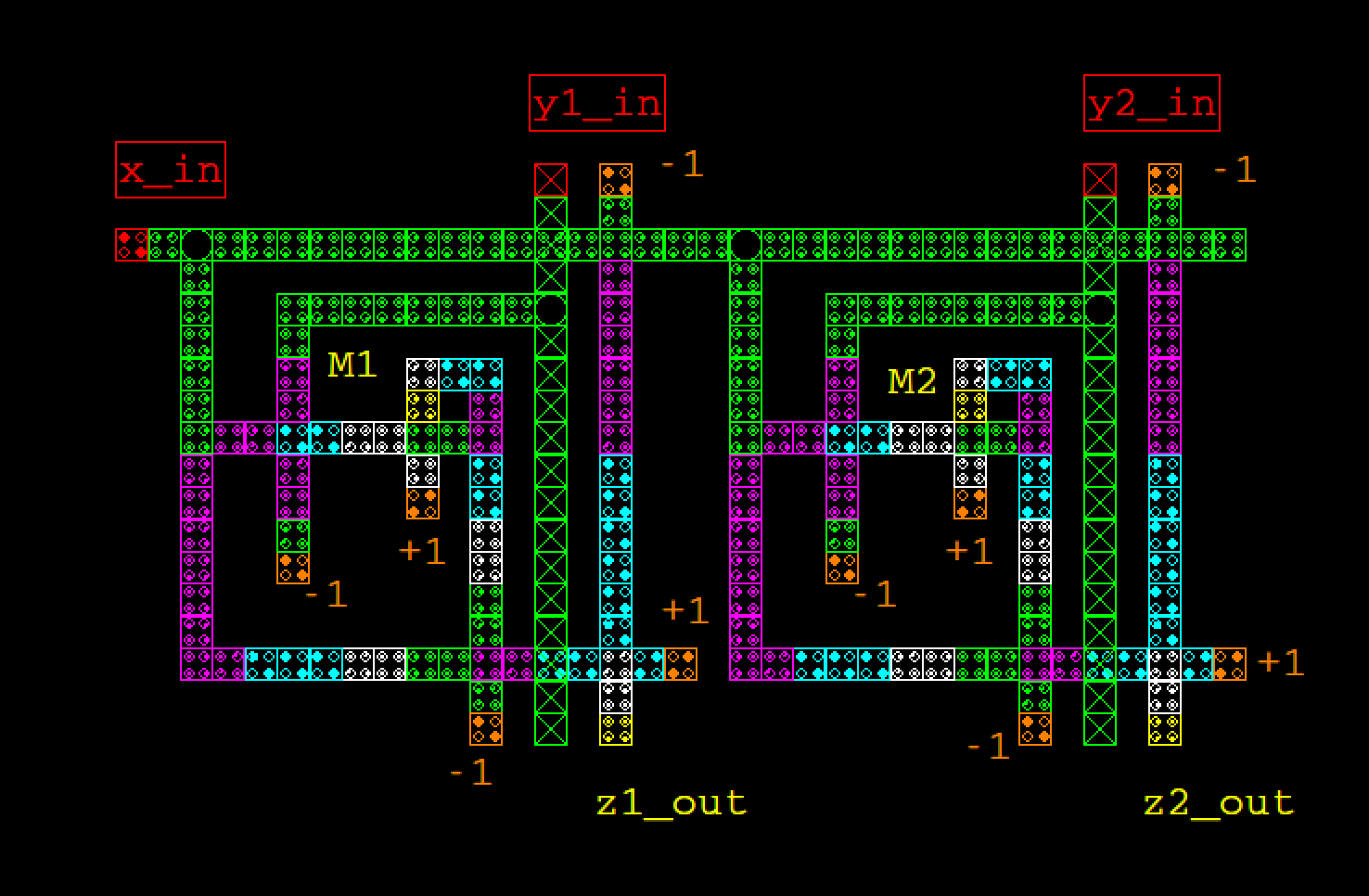} 
} %
} \\
 \subfloat[Vertical arrangement of two CMM neurons. As with the horizontal case in (a), the 
inputs are passed through the first neuron to the second, in this case the \texttt{y\_in} 
input is passed through, along with the \texttt{z\_out} outputs. The \texttt{z\_out} outputs 
need to be clocked through the design to preserve the output from each neuron, rather than 
all merging into a single bit output. \label{fig:two-neuron:design:vert}] {\centering
\parbox{0.8\linewidth} {  \centering
 \includegraphics[width=0.4\linewidth]{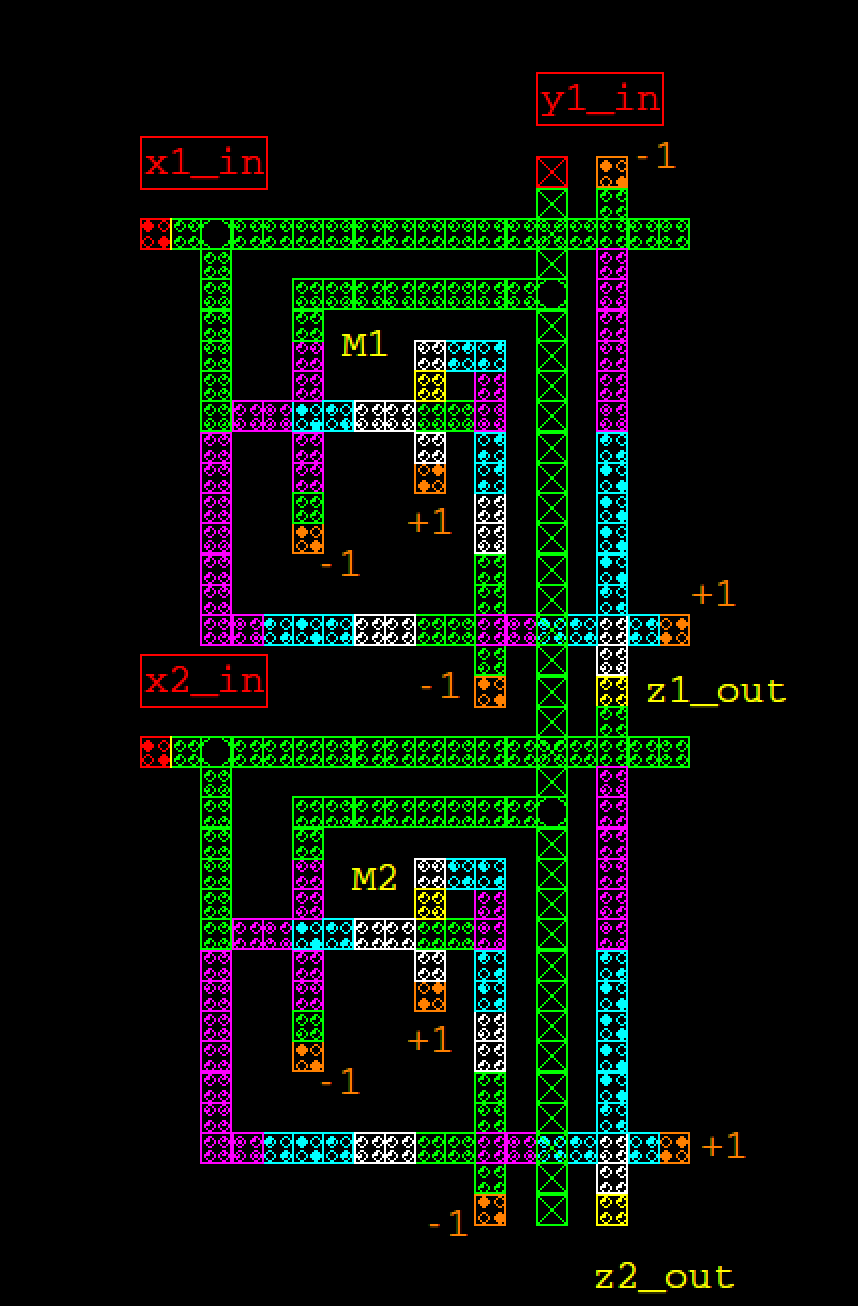} %
 } %
}
 \caption{}
 \label{fig:two-neuron:design}
\end{figure}

Fig.~\ref{fig:two-neuron:design} shows the selected approach to connecting multiple neurons 
together, in horizontal and vertical form. In this case, the problem of dealing with crossing 
lines of quantum-dot cells has been the focus, as the other problems only tend to appear in 
larger circuits. By keeping \texttt{x\_in} and \texttt{y\_in} on different layers to the main 
neuronal circuit, the design prevents these signals from interfering with the learning 
behaviour, and allows the same signal to be passed to both neurons within the same clock 
cycle.

%\clearpage{}

\begin{figure}[h!]
 \centering
 \includegraphics[width=0.7\linewidth]{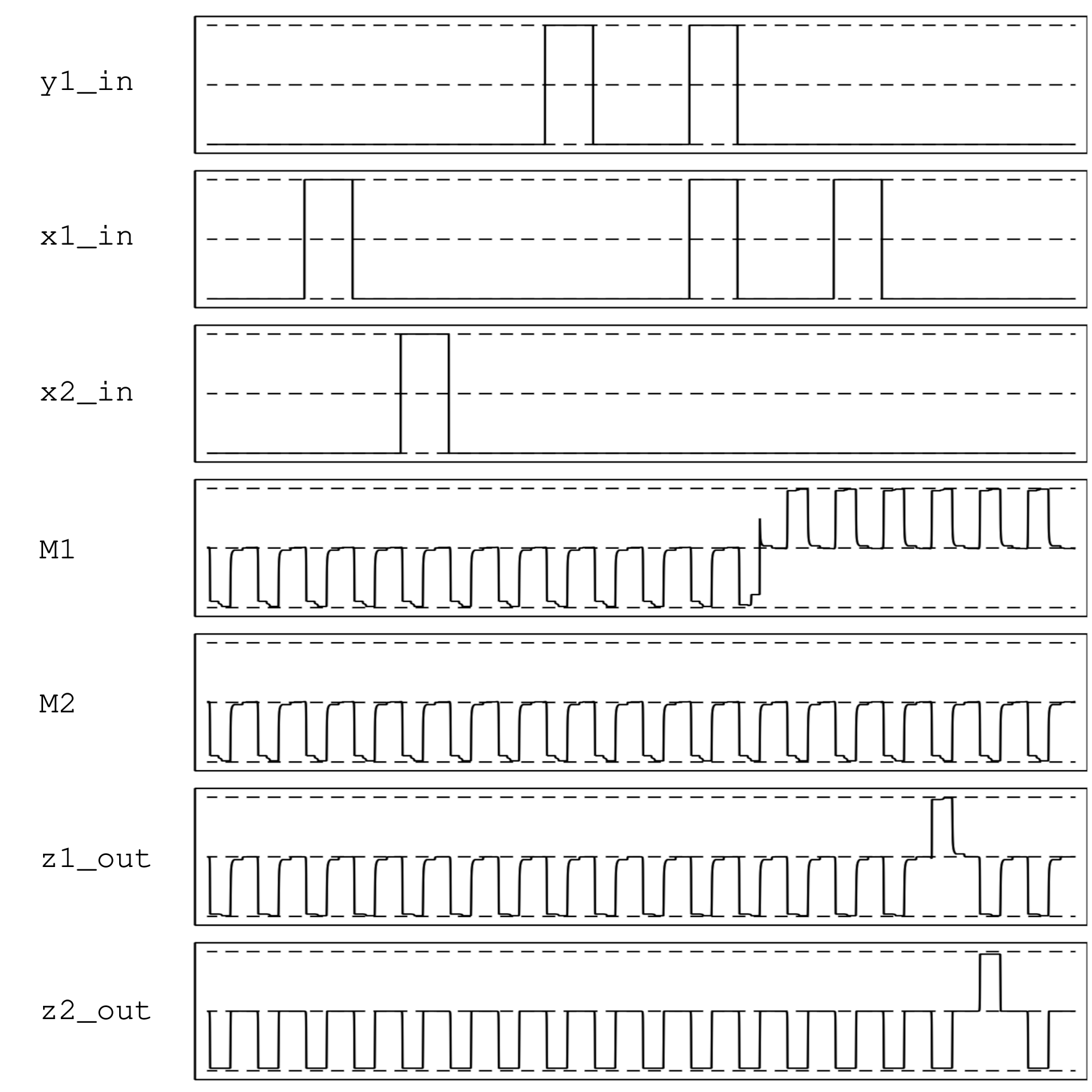}
 \caption[]{Simulation results for two vertically-arranged CMM neurons. The neurons are 
trained through two \texttt{x\_in} inputs and a single \texttt{y\_in} input, representing two 
bits in the binary input vector $\mathcal{I}$ and one bit in the binary output vector 
$\mathcal{O}$ (see sect.~\ref{sec:cmm}). The output from the first neuron \texttt{z1\_out} is 
passed through the same line of QCA cells as \texttt{z2\_out} to prevent the number of cells 
increasing super-linearly for each extra neuron. In order to prevent these output signals 
interfering, the outputs are clocked between each neuron, preserving the output from each 
neuron. As expected, \texttt{M1} is set when \texttt{x1\_in} and \texttt{y\_in} are logical 
1, and the value is recalled when \texttt{x\_in} is subsequently set to logical 1.}

 \label{fig:two-neuron:traces:vert}
\end{figure}

As with the single neuron testing phase, the behaviour of the neurons are tested using 
simulation to demonstrate that both horizontal and vertical arrangements of neurons train and 
recall as expected. Fig.~\ref{fig:two-neuron:traces:vert} shows the results of the 
simulations, where the vertical CMM array is trained based on the input bit vectors, and the 
information subsequently recalled from the network. Due to space constraints, the simulation 
results for the horizontal arrangement is provided in the appendix.

The \texttt{z2\_out} signal in fig.~\ref{fig:two-neuron:traces:vert} shows a logical 1 output 
immediately after \texttt{z1\_out} shows a logical 1 output as the signal is passed down 
through the same line of QCA cells. This does not signify that \texttt{z2\_out} is outputting 
a value, merely that the vertical combination of neurons is effectively using the QCA wire as 
intended.

%\clearpage{}

\subsection{4-neuron array}
 \label{sec:four-neuron}

 % connecting four neurons together (H)
 % issues encountered
 % - clocking
 % - interference
 % testing 4-neuron array
 
Unlike with the two-neuron case presented in section~\ref{sec:two-neuron}, problems were 
encountered when scaling to four neurons. The main issue was with the long lines of cells 
that act as wires in the QCA. Once they exceeded a certain length, they would not 
consistently communicate the signal from one end to the other.

This is likely a result of the clock speed being too high. As the propagation of state across 
the QCA cells takes time (around 2ps/cell~\cite{tougaw_dynamicbehaviorquantum}), the clock 
signal may change before the signal has reached the end of the QCA wire.\footnote{It is 
possible this is an artefact of the simulation, rather than a real-world problem, as it is 
expected that the clock signal will also experience propagation delay} To alleviate this 
problem, a clock cycle was introduced between each neuron on the \texttt{x\_in} wire that 
meant the signal was only required to transmit over a smaller distance within each clock 
cycle. In order to make sure the signals reach the neurons on the same clock cycles, a delay 
was also introduced on the \texttt{y\_in} inputs.

\begin{figure}[h!]
 \centering 
 \includegraphics[width=0.8\linewidth]{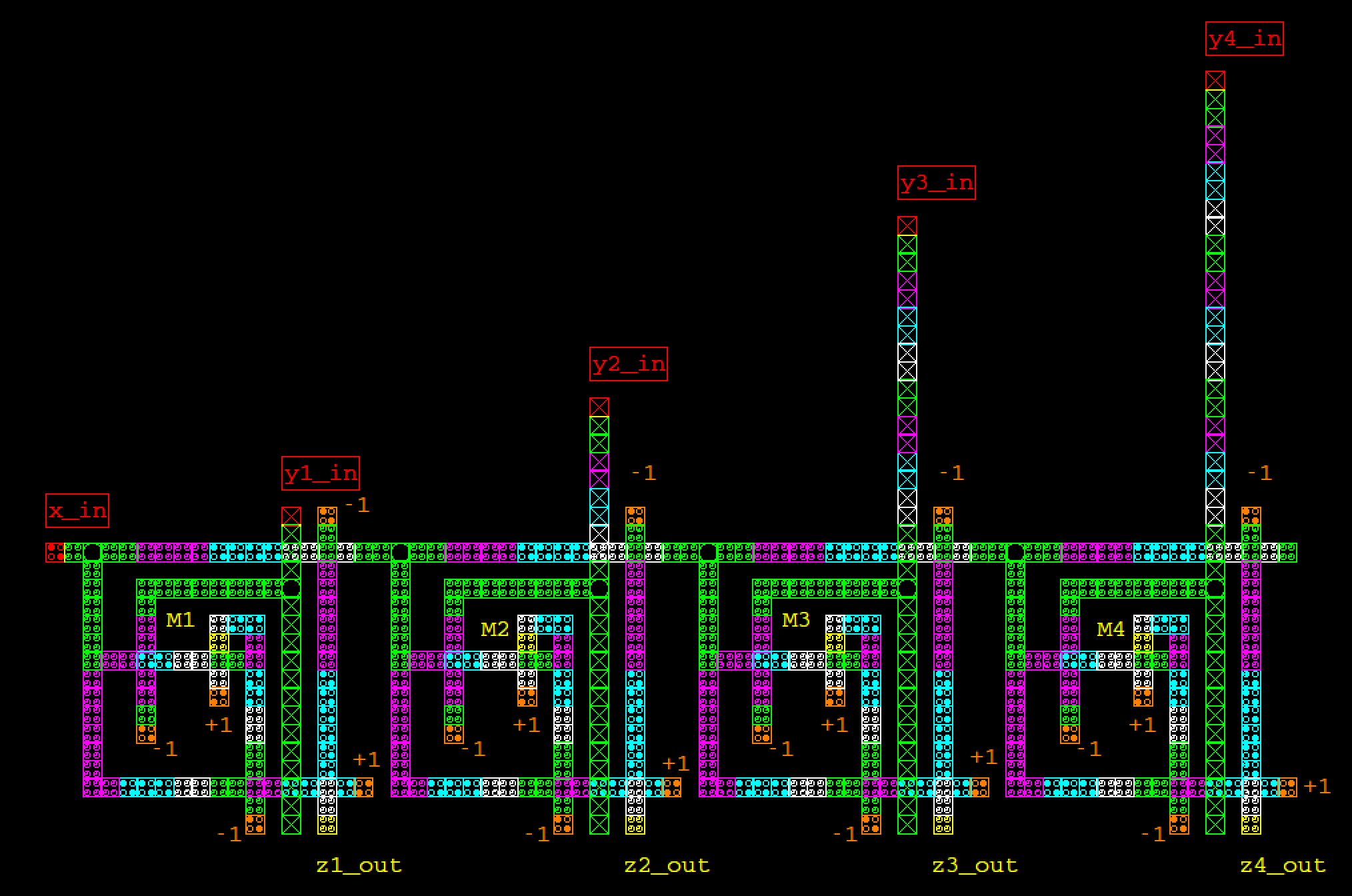}
 \caption[]{Four CMM neurons in a horizontal arrangement. The clock cycle required for 
long-range transmission is signified by the colours on the \texttt{x\_in} wire.} 
 \label{fig:four-neuron:design}
\end{figure}

Once the clocking mechanism was introduced into the CMM, four-neuron arrays were successfully 
implemented (see fig.~\ref{fig:four-neuron:design}). One of the negative side-effects of 
using this clocking method is that some of the parallelism that is so attractive for this 
substrate is lost---by using clocking, large arrays of neurons will take longer to train, 
changing from O(1) to O(n), where n is the number of clock cycles. If this problem is solved, 
this would prove to be an efficient method for training associative memory neural networks in 
QCAs. Due to space constraints, the simulation results for the four-neuron case are available 
in appendix~\ref{app:four-neuron:trace}.

\section{Discussion} 
\label{sec:discussion}

 % implementation efficiency
 % future work

 % neuronal density
 % rdc link (propagating kink across lines)

This is the first implementation of associative memory in quantum-dot cellular automata. 
Given the pattern-matching abilities of CMMs, and the ability to generalise from noisy 
inputs, this design could place CMMs at the forefront of QCA applications, once the 
fabrication of QCAs becomes efficient. The presented design has the potential to store circa 
$28\text{GB/cm}^2$ ($415\text{nm}^2$ per bit), or $240.5\text{ billion neurons/cm}^2$, 
multiple orders of magnitude higher than the density of neurons in the human 
brain~\cite{lange_cellnumbercell}.

The main limitation with the current proposed design is the need to introduce a clock delay 
on larger neuron arrays (see section~\ref{sec:four-neuron}). This drastically reduces the 
speed with which the CMM can be trained (from O(1) to O(n)). Removing this extra clock delay 
is currently the subject of further study.

Furthermore, the next logical step in developing the CMM is to introduce larger arrays of 
neurons, and design a threshold circuit for the output of the network which will enable the 
CMM to generalise noisy inputs (see sec.~\ref{sec:cmm}).

There is huge potential for using QCAs to study diffusive computation. In particular, using 
the propagation of state across the cells---the so-called 
`kink'~\cite{tougaw_dynamicbehaviorquantum}---as analogous to the propagation of waves in 
reaction--diffusion chemistry could be an interesting route forward for diffusive computing. 
It is possible that computation could occur by colliding these propagating kinks, reducing 
the reliance on clock cycles within QCAs and increasing the processing speed of QCA devices, 
while also providing a route to mainstream production for algorithms developed in the 
reaction--diffusion paradigm.

 % this is the first implementation of associative memory in QCAs
 % if used as basic RAM rather than associative memory, this design has the potential to 
 % store ~28GB/cm^2 (415nm^2 per bit)

 % CMMs are really good for pattern matching, and this design could improve the ability of 
 % CMMs to be applied in high-throughput and high-density environments

 % future work consists of dealing with the O(1) / O(n) problem,
 % thresholding and full arrays

\begin{comment}

  intro
  background on quantum dot cellular automata
  correlation matrix memories
  our implementation of CMMs in QCA
   -- individual neuron
   -- array of neurons (2x1, 4x1)
   -- training tests
  conclusion and next steps
  refs

\end{comment}

\clearpage{}

\appendix
\section{Supplementary simulation results}
\label{app:four-neuron:trace}

\subsection{Two-neuron (horizontal) simulation results}
\begin{figure}[h!]
 \centering
 \includegraphics[width=\linewidth]{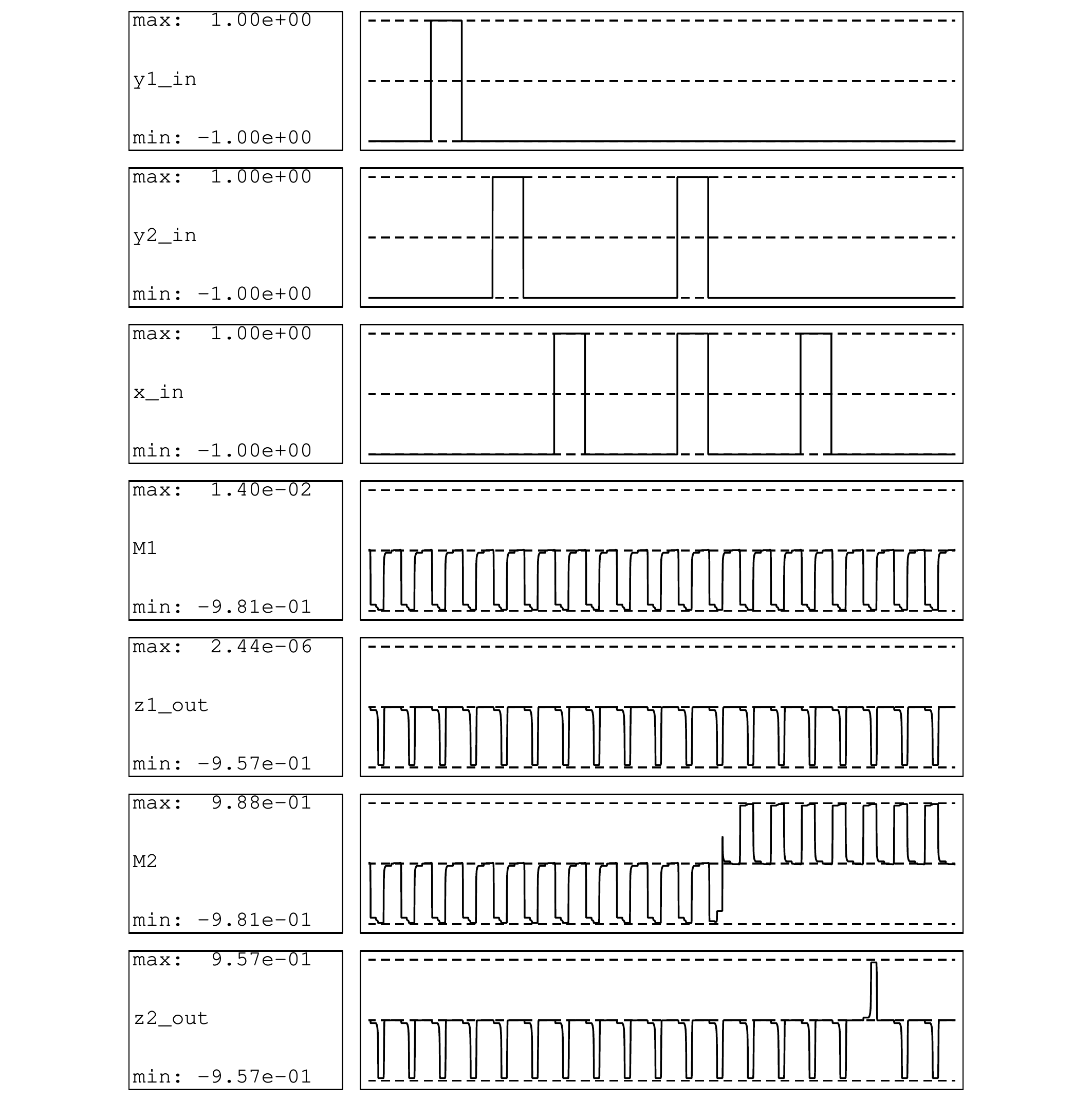}
 \caption[]{Simulation results for two horizontally-arranged CMM neurons. The neurons are 
trained through one \texttt{x\_in} input and two \texttt{y\_in} inputs, representing one bit 
in the binary input vector $\mathcal{I}$ and two bits in the binary output vector 
$\mathcal{O}$ (see sect.~\ref{sec:cmm}). As expected, \texttt{M2} is set when \texttt{x\_in} 
and \texttt{y2\_in} are logical 1, and the value is recalled when \texttt{x\_in} is 
subsequently set to logical 1.}
 \label{fig:two-neuron:traces:hoz}
\end{figure}
\clearpage{}

\subsection{Four-neuron simulation results}
\begin{figure}[h!]
 \centering
 \includegraphics[width=\linewidth]{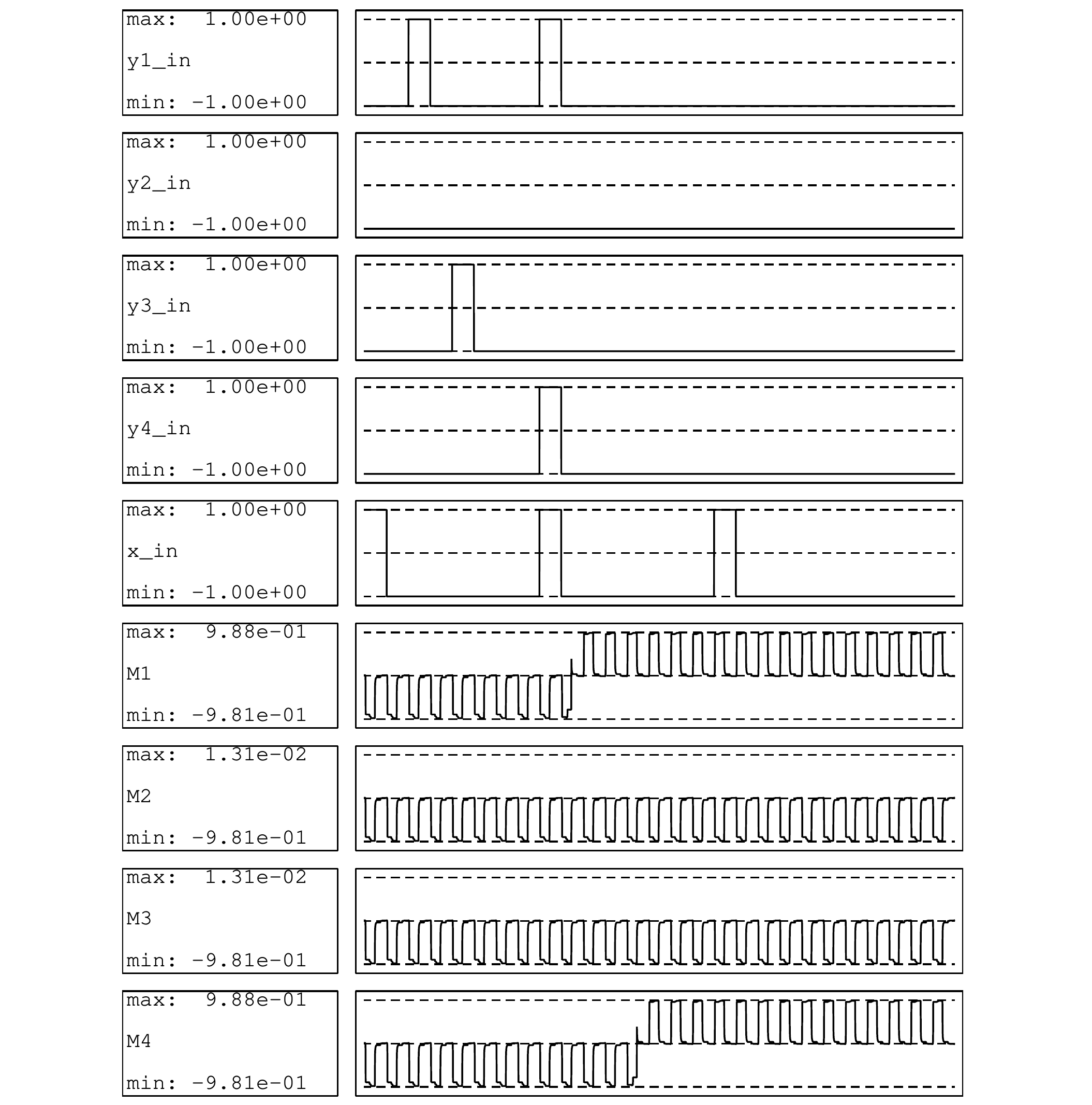}
 \caption[]{Simulation results for four horizontally-arranged CMM neurons. This figure 
presents the memory cell values, whereas fig.~\ref{fig:four-neuron:trace:recall} shows the 
output values. As expected, when trained with the binary vector $y = 1001$, the memory cells 
\texttt{M1} and \texttt{M4} are set to 1, the others remain as 0.}
 \label{fig:four-neuron:trace:store}
\end{figure}
\begin{figure}[h!]
 \centering
 \includegraphics[width=\linewidth]{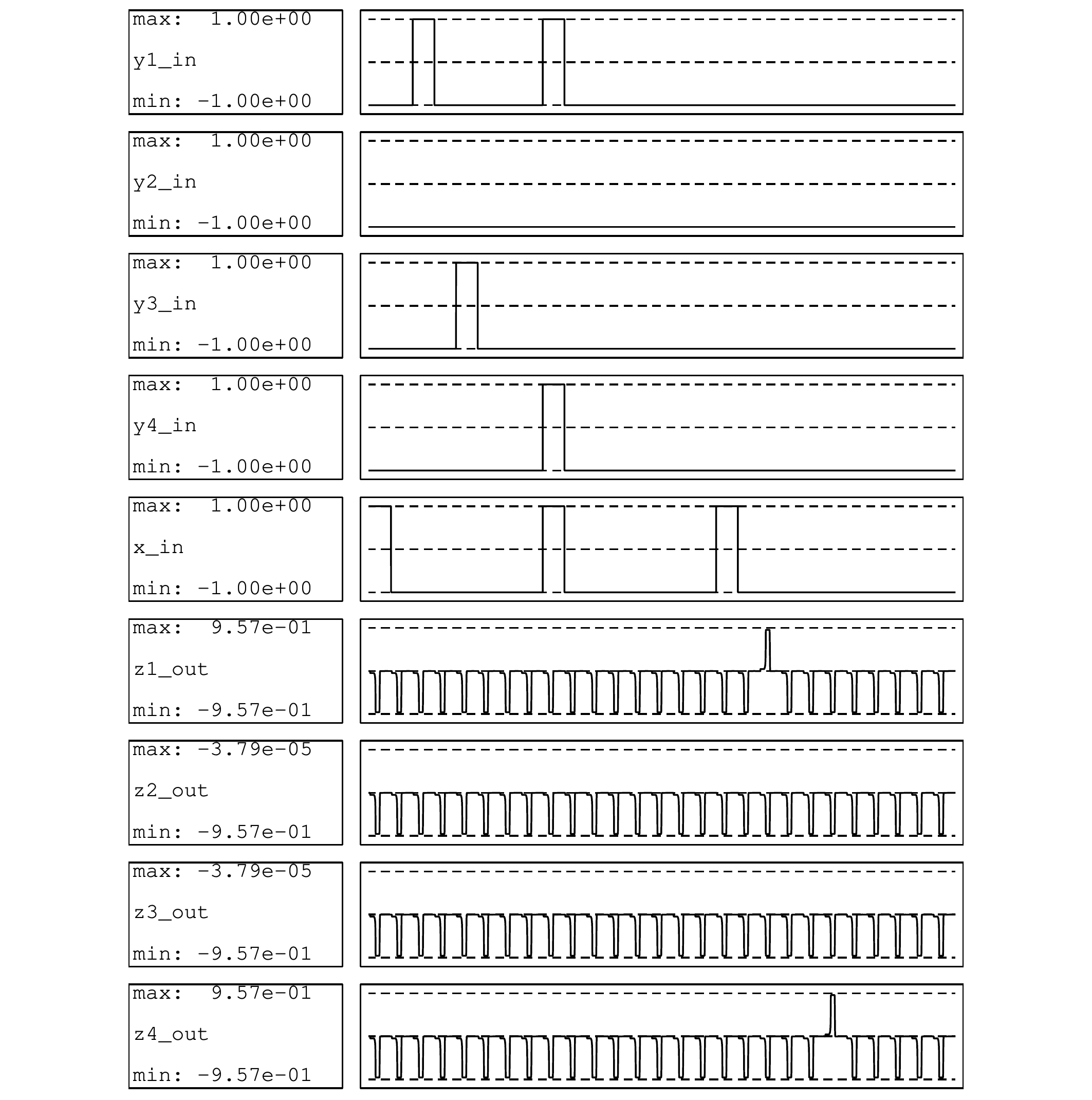} 
 \caption[]{Simulation results for four horizontally-arranged CMM neurons. This figure 
presents the recall results, whereas fig.~\ref{fig:four-neuron:trace:store} presents the 
training results. As expected, after training with binary vector $y = 1001$, the outputs from 
neurons 1 and 4 (\texttt{z1\_out} and \texttt{z4\_out}) are logical 1 upon recall and the 
others remain at 0.}

 \label{fig:four-neuron:trace:recall}
\end{figure}

\end{document}